Research Article

# Atomic density distributions in proteins: structural and functional implications

Sotirios Touliopoulos & Nicholas M. Glykos*

*Department of Molecular Biology and Genetics, Democritus University of Thrace, University campus, 68100 Alexandroupolis, Greece, Tel +30-25510-30620, Fax +30-25510-30620, https://utopia.duth.gr/glykos/ , glykos@mbg.duth.gr*

# Abstract

Atomic packing is an important metric for characterizing protein structures, as it significantly influences various features including the stability, the rate of evolution and the functional roles of proteins. Packing in protein structures is a measure of the overall proximity between the proteins' atoms and it can vary notably among different structures. However, even single domain proteins do not exhibit uniform packing throughout their structure. Protein cores in the interior tend to be more tightly packed compared to the protein surface and the presence of cavities and voids can disrupt that internal tight packing too.

Many different methods have been used to measure the quality of packing in proteins, identify factors that influence it, and their possible implications. In this work, we examine atomic density distributions derived from 21,255 non-redundant protein structures and show that statistically significant differences between those distributions are present. The biomolecular assembly unit was chosen as a representative for these structures. Addition of hydrogen atoms and solvation was also performed to emulate a faithful representation of the structures *in vitro*.

Several protein structures deviate significantly and systematically from the average packing behavior. Hierarchical clustering indicated that there are groups of structures with similar atomic density distributions. Search for common features and patterns in these clusters showed that some of them include proteins with characteristic structures such as coiled-coils and cytochromes. Certain classification families such as hydrolases and transferases have also a preference to appear more frequently in dense and loosely-packed clusters respectively.

Regarding factors influencing packing, our results support knowledge that larger structures have a smaller range in their density values, but tend to be more loosely packed, compared to smaller proteins. We also used indicators, like crystallographic water molecules abundance and B-factors as estimates of the stability of the structures to reveal its relationship with packing.

# 1. Introduction

Atomic packing has been an important metric for characterizing protein structures since 1974, when it was observed that the average packing density within proteins' interiors is roughly equivalent to that of small organic molecule crystals [1]. Although numerous methods had been developed to calculate the packing and interactions of amino acid residues within proteins, the use of packing density as a criterion for evaluating model protein structures was developed explicitly in 1990 [2].

To date, several approaches have been tested to measure the atomic packing in structures. The Voronoi procedure is a widely-used method, in which a unique volume is assigned to individual atoms in order to study variations in packing of proteins [3–7]. Another well-established method for analyzing packing interactions in proteins is based on the calculation of the occluded molecular surface [8]. Other methods and approaches for analyzing protein packing have also been reported [9–14].

Packing is an important aspect of protein structures. A compact packing of amino acid residues is known to affect both the thermal stability and folding rate of proteins. [15–24]. Protein stability is of significant interest to the biotechnological, pharmaceutical, and food industries. The effects of packing on protein stability has extensively been studied to the point that modeling programs have incorporated packing as a parameter, aiming to predict protein stability after mutations [25]. Moreover, hydrogen bonds, which increase the packing density in the protein interior are known for their indirect contribution to protein stability [26]. Recent studies indicated that the major determinants of protein stability include packing and van der Waals interactions [27–29].

As the structural comparison shown in Fig.1 and Fig.2 exemplifies, proteins do not exhibit uniform packing throughout their structure [30]. Localized packing defects appear as cavities, and their presence can compromise the stability of the protein [31]. Additionally, the distribution of these voids (cavities) is highly heterogeneous across different proteins [32]. Despite the presence of occasional cavities, the interior of spherical proteins remains tightly packed. The Voronoi volumes of surface atoms, modeled with solvent surrounding the protein, are approximately 7% larger [33,34], indicating that packing is less dense on the protein surface.

Experimental studies have shown that mutations in protein cores, where small residues are replaced with larger ones, generally destabilize the protein. This suggests that there is minimal empty space available to accommodate additional atoms [22,35]. This can be explained by the α-helical and β-sheet secondary structures in globular proteins. These elements organize in a manner that allows non-polar side chains to interlock like jigsaw puzzle pieces, creating densely packed cores. As a result of this tight packing, van der Waals forces are considerably stronger in the interior [26]. However, due to energetically unfavorable atomic overlaps, protein cores cannot exceed some density limits [36]. The rigidity of protein cores is also shown to be strongly correlated with packing density [37,38]. Furthermore, studies have shown that the interior of proteins evolves slowly, in contrast to the surface which has more rapid evolution [39,40]. Solvent accessibility has become the *de facto* structural measurement to use in protein evolution studies. However, more recent work has called the central role of solvent accessibility into question and has identified packing as an important factor too [41]. The two packing measures most frequently employed in evolutionary studies are the contact number and the weighted contact number. For a given amino acid, the contact number represents the total count of other residues within its local structural neighborhood. In contrast, the weighted contact number considers all residues in the protein, assigning weights to them based on the square of their inverse distance to the amino acid under examination [42,43].

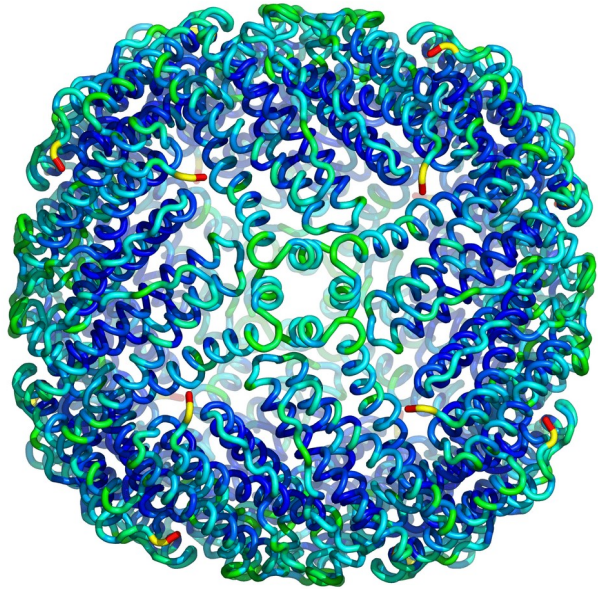

**Figure 1:** Schematic diagram of a ferritin-like protein structure with a low density profile. This is the biological assembly corresponding to PDB entry 3r2k colored according to atomic temperature factors.

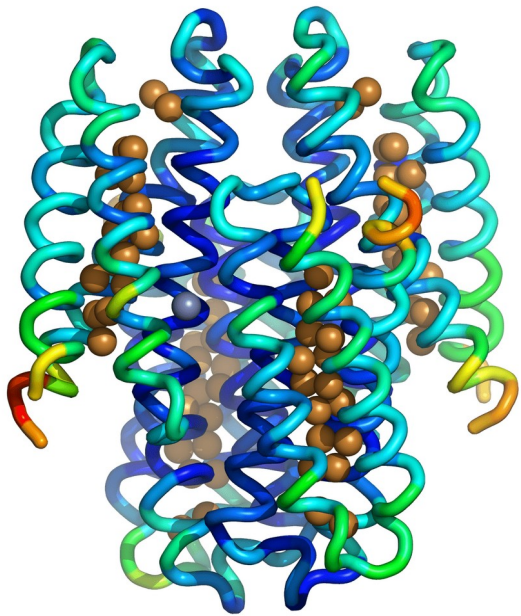

**Figure 2:** Schematic diagram of the structure of a protein with a high density profile. This is a cytosolic copper storage protein (PDB entry 6zif) colored according to atomic temperature factors, with the copper atoms indicated as solid spheres.

Atomic packing is affected by a combination of different factors. A statistical analysis of the radius of gyration for 3,769 protein domains across four major classes (α, β, α/β, and α+β) revealed that each class exhibits a characteristic radius of gyration, indicating its specific level of structural compactness. For example, α-helical proteins exhibit the highest radius of gyration across the considered protein size range, indicating a less compact packing compared to β and (α + β) proteins. In contrast, α/β proteins display the lowest radius of gyration, characteristic of the most compact packing among the classes [44].

Another study showed that for proteins with a molecular weight below 20 kDa, the average density shows a positive deviation that becomes more pronounced as molecular weight decreases, indicating that smaller proteins are more densely packed than larger ones, which tend to have a looser packing structure [45]. Additionally, an analysis of 152 non-homologous proteins demonstrated that variations in protein packing are influenced by a complex interplay of protein size, secondary structure, and amino acid composition. They showed that helices appear to be more efficiently packed compared to strands and that large proteins are expected to have increased overall packing[46].

In this communication we attempt to approach the problem of characterizing and analyzing protein density not through average statistical or structural properties, but by building and directly comparing individual density profiles which were created for an extended set of more than 21,000 proteins. The essence of our approach is the following : For each atom of each structure we calculate the density (in Da/Å$^3$) inside a sphere centered on that atom. If, for example, a given protein structure contains 5,000 atoms, then we would calculate 5,000 density values (one for each atom). These density values are then used to calculate a histogram of their distribution which is characteristic of the protein structure under examination. Having collected the density distributions, we can quantify and analyze their similarities and differences using established metrics such as the Euclidean distance (calculated between any given pair of distributions). By doing an all-to-all comparison of those distributions, we can quantitatively characterize structural and functional patterns present in these distributions. In the following paragraphs we present details of this method, and of the structural and functional results obtained from its application.

## 2. Methods

### 2.1 Calculation of density distributions

The starting set of protein structures (see previous paragraph) is an extended sample obtained from the PDB and comprising representative culled proteins (identity cutoff <= 50%), hydrogenated and hydrated to simulate *in vitro* conditions (see section §2.2 below for details of how the representative structures were obtained and how hydrogens and waters were added). Throughout our calculations we have tested three different sphere radii (5Å, 6Å, 7Å) to remove the bias that this otherwise arbitrary choice would incur. Greater values (> 7Å) have also been tested but were found lacking the resolution needed by our method. In the first stage, our method parses the *x,y,z* coordinates of atoms and calculates the distance between every possible combination of atoms (*i, j*). For any given radius R (5Å, 6Å or 7Å), it calculates whether the distance between the atoms *i* and *j* is smaller than R, and if so, the mass (in Dalton) of atom *j* is counted as lying inside the sphere of atom *i*. This is performed recursively for every atom, and after division with the volume of the sphere, the atomic density distribution of the protein under examination is obtained (in units of Dalton per cubic Ångström). Note that this approach is based on counting only the presence or absence of atoms lying inside spheres centered on every other atom. Each atom contributes as a whole to the density calculation when it lies inside a sphere, with no correction being made for the part of the atom's volume that is outside the sphere. Finally, we note that this algorithm is only applied to protein atoms (and not, for example, to the water molecules that were added to simulate a fully hydrated protein structure). Fig.3 shows a detailed flow chart of this procedure.

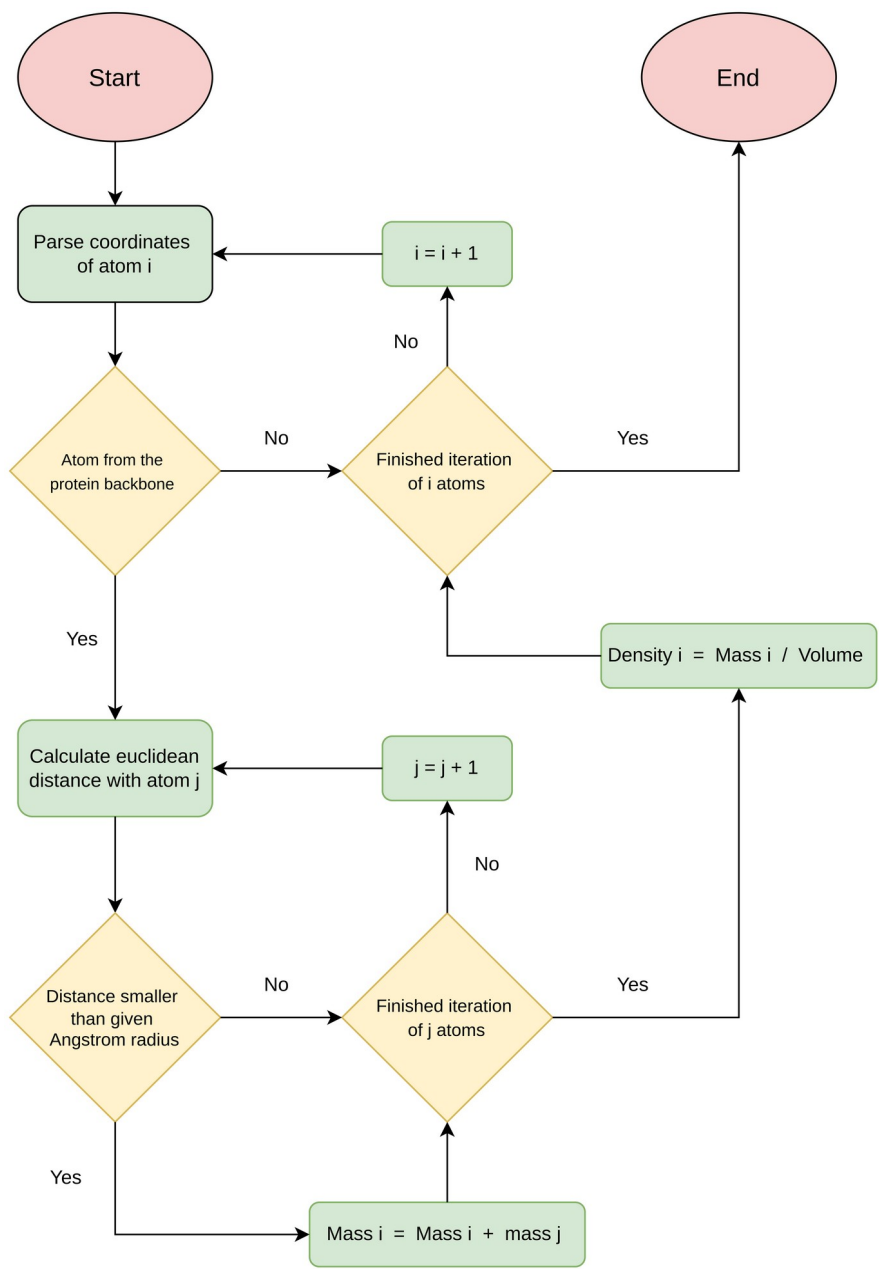

**Figure 3:** Flow Chart for the calculation of the density profiles.

## 2.2 Representative structures and system preparation

The PISCES server[47] was used to obtain a set of proteins with diverse structural and functional characteristics from the Protein Data Bank. The full list of selection criteria given to PISCES are shown in Table 1, which resulted to a set of 22478 structures. For all structures, the biological assembly was used for all further calculations. An additional cutoff of a maximum of 80,000 protein atoms per structure was applied to final list of structures.

**Table 1:** Criteria given to PISCES server for culling the Protein Data Bank

| Criteria | Value |
| --- | --- |
| Resolution | 0.0 - 2.2 |
| R-factor | 0.25 |
| Sequence length | 50 - 10000 |
| Sequence percentage identity | <= 50.0 |
| X-ray entries | Included |
| EM entries | Excluded |
| NMR entries | Excluded |
| Chains with chain breaks Included | Included |
| Chains with disorder | Included |

OpenBabel[48] was used to add missing hydrogen atoms to PDB files. The program *Solvate* (https://www.mpinat.mpg.de/grubmueller/solvate) was used to perform hydration of structures by adding and energy minimizing a sufficiently large box of pre-equilibrated water molecules around the solute, emulating a fully hydrated protein structure *in vitro*. Care was taken to avoid adding water molecules in buried protein cavities. Figure 4 shows a schematic illustration of this procedure for the case of a small protein.

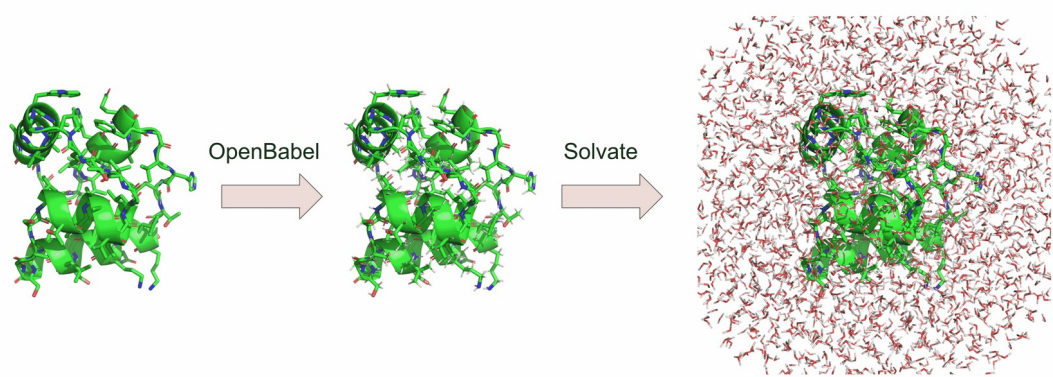

**Figure 4:** Preparation of the final systems before density calculation : Addition of hydrogens using *OpenBabel*, and of a pre-equilibrated water box using *Solvate*.

# 3. Results

**3.1 Cumulative distributions indicate the presence of significant density variability**

The density distributions (one for each protein examined) were calculated as described in section §2.1. To simplify the subsequent calculation of distance metrics between different distributions, the same number of histogram bins (equal to 100) was used for all proteins. This choice for the number of bins was guided by the application of the Freedman-Diaconis rule to a randomly selected subset of protein structures. In the final step of preparing the initial data set, the individual distributions were normalized by dividing with the total number of atoms of each protein, thus converting the units to frequencies (of observing the corresponding density, see Fig.5 below). The final data set comprises three (21255x100) matrices, corresponding to the three radii we examined (5Å, 6Å, 7Å). In each of these matrices, every row corresponds to a different structure and every column to a different bin from the density distribution of the given protein (and for the given radius). Fig.5 shows the cumulative density distributions (i.e., over all 21255 proteins) for each of the radii we examined.

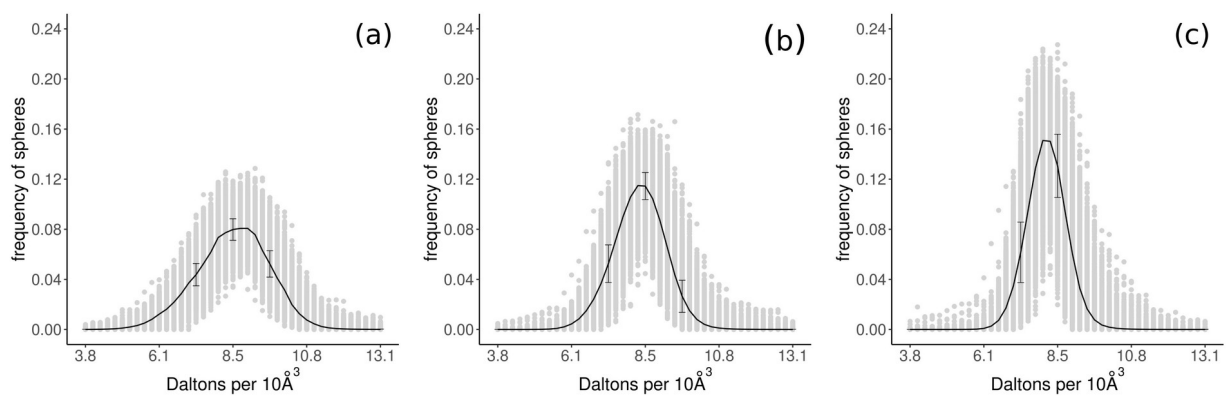

**Figure 5:** Cumulative density distributions for each of the radii examined : (a) is for the 5Å radius, (b) 6Å, (c) 7Å. In each diagram, the black line corresponds to the mean value of the respective bins. A representative sample of the corresponding standard deviations has also been added to the diagrams to help establish the amount of variance present in the individual distributions.

It can be seen that several data points are significantly distant from the mean, with several of them deviating by more than 3σ from the average. This implies that there could be certain distributions that would be characterized as 'outliers' when compared with the average distribution. The observation that as the radius increases, the distributions become tighter about their mean is fully consistent with our expectations : as the volume of the spheres increases, the density calculated from each sphere approaches the same value (which is the average density of protein structures). Equivalently, as the radius is increasing, the 'high resolution' information about the variation of density inside a protein structure is diminished due to extensive averaging.

## 3.2 Principal Component Analysis allows the identification of unusual density distributions

Before proceeding with the main theme of our analysis –which is based on the clustering of the primary data through hierarchical clustering methods (discussed below)– we used PCA as a preliminary step to visualize the distribution of data in the reduced principal component space. Figures 6(a,b,c) show the PCA distributions obtained from the matrix calculated with the 6Å radius and projected on the top three principal components. We observe that the data points are not harmonically distributed in these projections. There is a pronounced high density area (corresponding to an 'average' density distribution), but significant excursions from normality are immediately obvious (for example the tails clearly seen in Figure 6(a)). To help visualize the amount of deviation present in the density distributions, we show in Figure 6(d) the average density distribution (black line) *versus* the density distributions obtained from two outliers (structures 1J0P and 3NIO). The 1J0P structure is a cytochrome and its density distribution is markedly shifted to the right (higher density values). The 3NIO structure on the other hand is shifted to the left (lower density values) and corresponds to a guanidinobutyrase protein. This is an early indication that structures with uncommon distributions exist. Further examination of outliers reveals structures with a preference in light-harvesting (cytochromes) and copper-storage proteins. The biological significance of these deviations from a harmonic behavior are analyzed in the following sections.

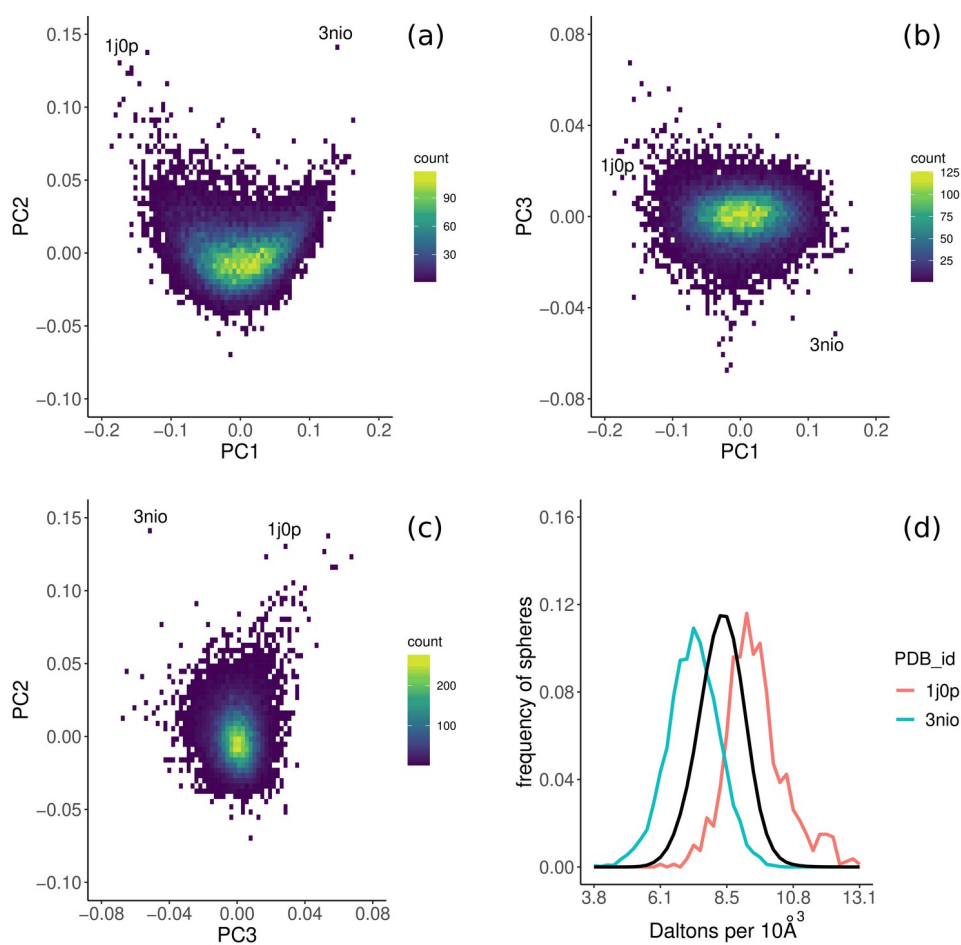

**Figure 6:** Principal Component Analysis of the matrix obtained with a 6Å radius. See text for details of this analysis.

**3.3 Hierarchical clustering allows the identification of distinct groups of proteins**

Starting from the three (21255x100) matrices corresponding to the three radii we examined (see section §3.1), we calculated the corresponding distance matrices by calculating the Euclidean distance between all possible pairs of distributions. These distance matrices are the primary data upon which hierarchical clustering methods are based. Figure 7 shows a visualization of these symmetrical distance matrices in the form of heatmaps where distances are encoded as colors ranging for dark blue (small distances), through yellow (intermediate distances), to red (large distances). The lower three panels in Fig.7 show the same matrices but after exclusion of the 100 most distant proteins (this was done in order to increase the dynamic range of these graphs). Please note that all matrices shown in Fig.7 have been scaled to the same maximum distance (they are on the same color scale).

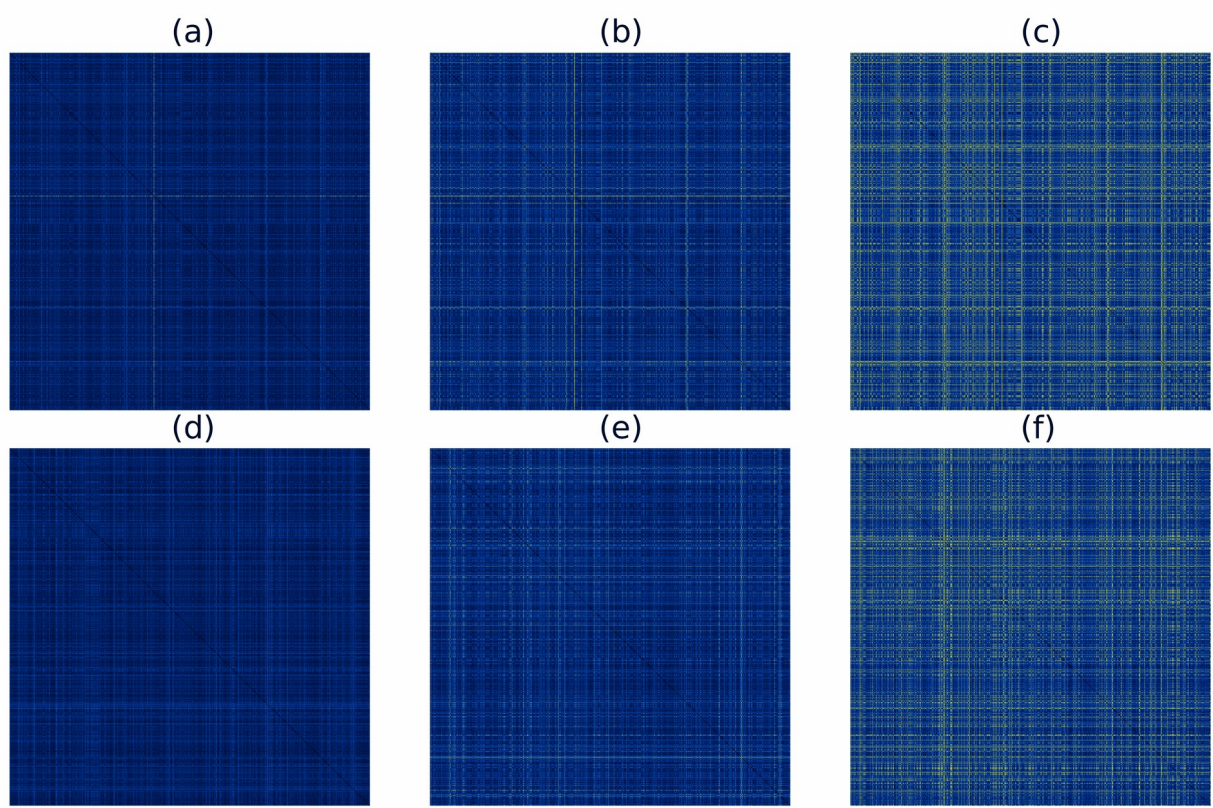

**Figure 7:** Distance matrices visualized in the form of heatmaps. Panels (a), (b) and (c) correspond to the complete 5Å, 6Å, and 7Å matrices. The lower three panels (d,e,f) are the same matrices but after removal of the 100 most distant structures to allow visualization of smaller distances in the matrices.

The general appearance of the distance matrices shown in Fig.7 is fully consistent with the results discussed in sections §3.1 and §3.2. For example, the increase of the average distances as we move from the 5Å matrix, to 6Å, and to 7Å (going from mostly blue colors to mostly yellows in Fig.7), is consistent with the tighter and higher density distributions seen in Fig.5 : Higher values of frequencies (as seen, for example, in Fig.5(c)) lead to larger on average Euclidean distances, leading to the systematic trend observed in Fig.7. To put this in numbers, Table 2 shows the averages, standard deviations and maximal distances recorded for the three distance matrices shown in Fig.7.

**Table 2:** Statistics for the three distance matrices shown in Fig.7.

| Matrix | Average distance | Standard deviation | Maximum distance |
|--------|------------------|--------------------|------------------|
| R=5Å   | 0.05             | 0.02               | 0.29             |
| R=6Å   | 0.07             | 0.04               | 0.35             |
| R=7Å   | 0.10             | 0.06               | 0.47             |

The second important feature of these matrices concerns their internal consistency (which, however, is more difficult to discern due to their relatively low contrast of the heatmaps). Closer examination of the graphs in Fig.7, however, does show that the patterns of small/large distances (dark/light colors) is more-or-less the same irrespectively of which matrix is being examined. To put this observation in numbers, we compared these three matrices by calculating the values of the linear correlation coefficient between all possible pairwise combinations. The comparison of the (5Å matrix) *vs.* (6Å matrix) gave a value of the linear correlation coefficient of +0.86, the 6Å *vs.* 7Å comparison gave a value of +0.93, and the 5Å-7Å combination a value of +0.76. The fact that the matrices are so similar and internally consistent is reassuring : it implies that the subsequent calculation of dendrograms is robust and not highly sensitive to the value of the averaging radius. The lower correlation for the 5Å-7Å pair, combined with the value of +0.93 for the 6Å-7Å pair indicates the information content of the matrices has stabilized (converged) once we reach the 6Å radius. For this reason, all further calculations reported in this communication were based on the matrix calculated with the 6Å averaging radius [diagrams (b) and (e) in Fig.7].

Figure 8(a) shows the dendrogram obtained by performing hierarchical clustering of the 6Å distance matrix using the R package for statistical computing. Hierarchical clustering produced well-separated clusters of structures, with the individual clusters having similar atomic density distributions and being of a size suitable for further statistical analysis. Other clustering algorithms such as k-means and HDBSCAN have also been tested, but failed to produce well-separated clusters (which is not unexpected given the relatively uniform distribution of the raw data as principal component analysis clearly indicated, see Fig.6). However, k-means and HDBSCAN did provide additional information on outliers or small groups of structures with common characteristics.

The fact that the dendrogram is well-structured does not alleviate the problem of how to select the cutoff distance ('height') needed for cluster definition. Using the data shown in Table 2, a cutoff of (mean+2σ) for the 6Å matrix would have given a value for the cutoff of 0.15 units. We have elected to slightly lower this number to 0.14 units in order to differentiate between the two clusters that their lineage separated at that height (these are the clusters shown in purple and cyan in Fig.8(b), the 2[nd] and 3[rd] from the left). This selection resulted to a total of 12 clusters. All further analyses discussed below will be referred to these clusters.

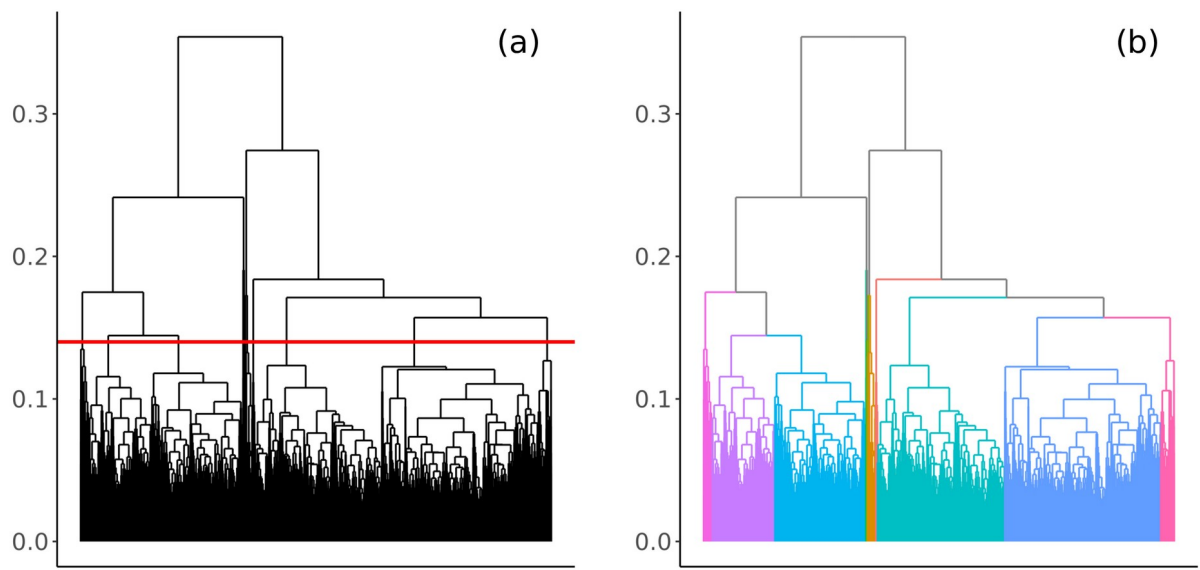

**Figure 8:** Hierarchical clustering. Panel (a) shows the dendrogram produced from the 6Å distance matrix using the "complete" linkage method. The red line corresponds to the height (*h*=0.14) where the dendrogram was cut to produce clusters, see text for details. Panel (b) is the same dendrogram but with the individual clusters color-coded to aid identification.

Examination of the dendrogram indicates that these 12 clusters vary significantly not only in size, but also on their distance separation (the 'height' of their last common ancestor). To be able to focus on the distant clusters (which are the most informative), we decided to quantify the distances between these 12 clusters by calculating a Z-score matrix (one Z-score for each possible pairwise cluster combination). To perform this calculation, we started by fitting the atomic density distributions of the proteins that belong to each cluster to a Gaussian (see Fig.9(b) for a pictorial explanation of the procedure). Once the means and standard deviations were available for each cluster, the Z-score matrix could be constructed, and used for another round of hierarchical clustering. The resulting dendrogram (this time showing relationships between *clusters* of proteins) is shown in Fig.9(a).

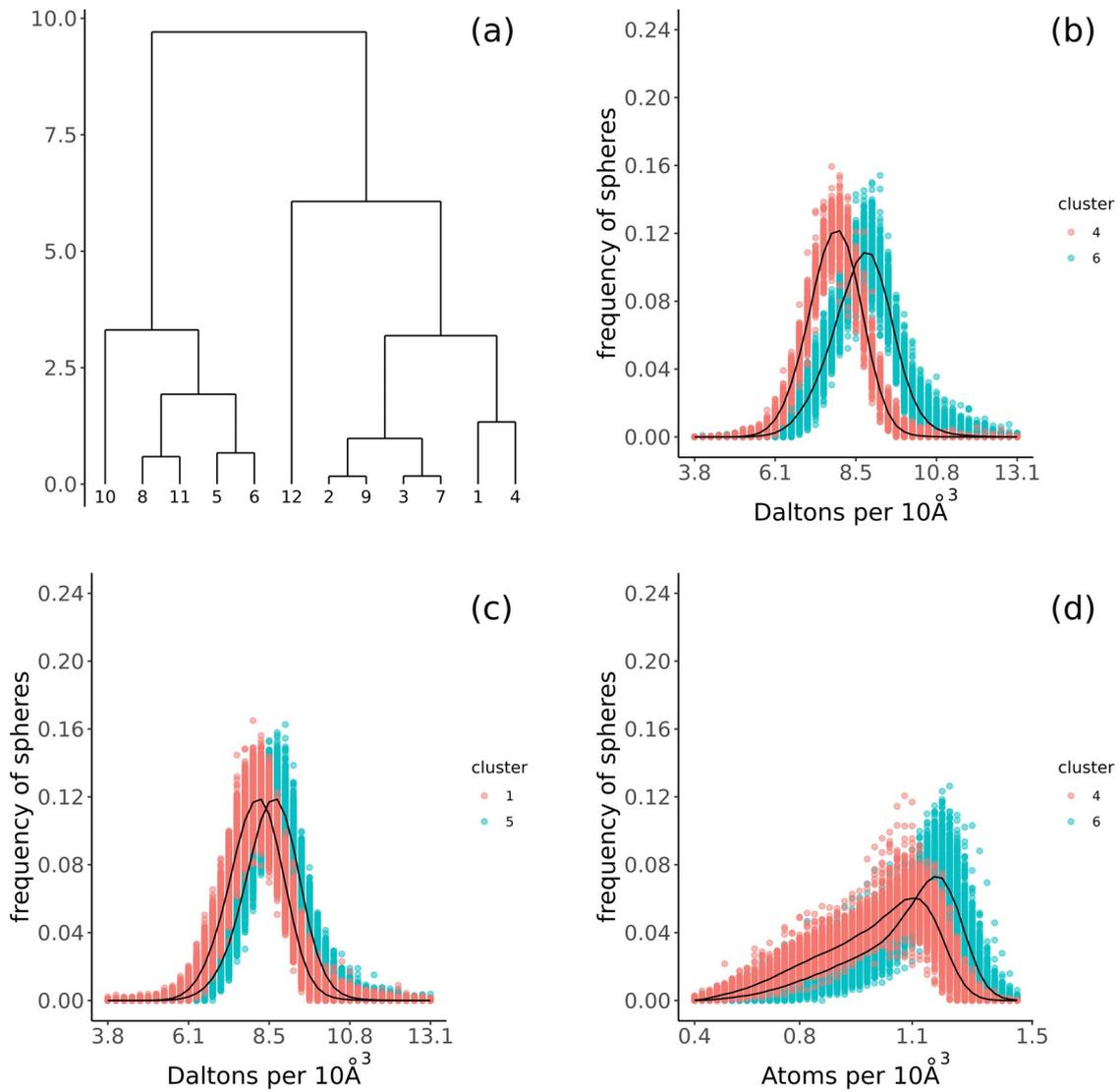

**Figure 9:** Groups of structures deviate significantly from the average behavior. (a) Dendrogram from hierarchical clustering of the Z-scores matrix using the "complete" linkage method. (b) Comparison of the raw data of the clusters 4 and 6. (c) Comparison of the raw data of the clusters 1 and 5. (d) Comparisons of the raw data of the clusters 4 and 6, but in units of atoms per cubic Ångström, see text for details.

The availability of this dendrogram, and after exclusion of clusters with less than 100 members, allowed us to focus on just five clusters of proteins that demonstrate significant deviations between them. Statistics for those five clusters are shown in Table 3. Figures 9(b) and 9(c) show an explicit (raw data-based) comparison between the density distributions of proteins that belong to clusters 1, 4, 5 and 6. These graphs clearly indicate the presence of significant deviations at the raw-data level, with two clusters showing a shift to lower densities (clusters 1 & 4), and two showing a shift to higher densities (clusters 5 & 6).

**Table 3:** Statistics for the five clusters of interest sorted on the basis of their median atomic density.

| Cluster | Sample size | Median density |
|---------|-------------|----------------|
| 4 | 344 | 0.77 |
| 1 | 5813 | 0.79 |
| 3 | 7045 | 0.81 |
| 5 | 2821 | 0.84 |
| 6 | 400 | 0.85 |

One important question at this point, is whether the differences of the distributions seen for example in Fig.9(b) are due to the presence of ligands containing heavy atoms, or whether they reflect genuine differences in the packing of atoms. To tackle this issue, we recalculated for clusters 4 & 6 the protein density distributions, but instead of using units of "Daltons per cubic Ångström", we calculated the distributions in units of "atoms per cubic Ångström", thus ignoring the atomic weights of the atoms involved. The results from this calculation are shown in Fig.9(d). Comparison between panels (b) and (d) in this figure indicates that a genuine difference of the atomic packing of the proteins that belong to these clusters appears to be present (and is not just systematic differences in the presence of ligands that lead to the observed differences). In the sections that follow we present some indications concerning the source of the observed systematic density deviations.

**3.4 Structural implications of the atomic density distributions**

In this section we examine the relationships between the density distributions we obtained above *versus* structural characteristics of the respective proteins such as size, secondary structure, temperature factors, and abundance of water molecules.

The first observation is that the number of residues is anti-correlated with the median atomic density, see Fig. 10. The Spearman correlation coefficient is equal to -0.18 for the 6Å data, and -0.27 for the 5Å radius data. This finding is in good agreement with data from other studies which indicated that larger proteins tend to be packed more loosely than smaller ones [50,51]. Examination of the scatter plot in Figure 10(b), indicates that as the number of amino acids increases, the variance of the median density values decreases. This finding may indicate the presence of systematic structural/thermodynamic or evolutionary constraints that make a tight atomic packing of larger proteins uncommon. To put this observation in numbers, we divided the sample based on the quartiles of the residues. The median and standard deviation values of each quartile are shown in Table 4. There is a pattern in both metrics when going from samples representing small proteins to samples representing large structures, with both median and standard deviation decreasing in larger structures. Small proteins seem to have increased variability when it comes to packing, as indicated by their wider density limits. This finding may align well with the conclusion of a previous study that "Proteins are not optimized by evolution to eliminate packing voids" [51]. It is also in good agreement with an additional study, which showed a reduction in the range of density values in larger proteins. In their analysis small proteins exhibit a broad range of packing densities, varying from 0.67 to 0.87, while for large proteins densities range from 0.69 to 0.74 [52].

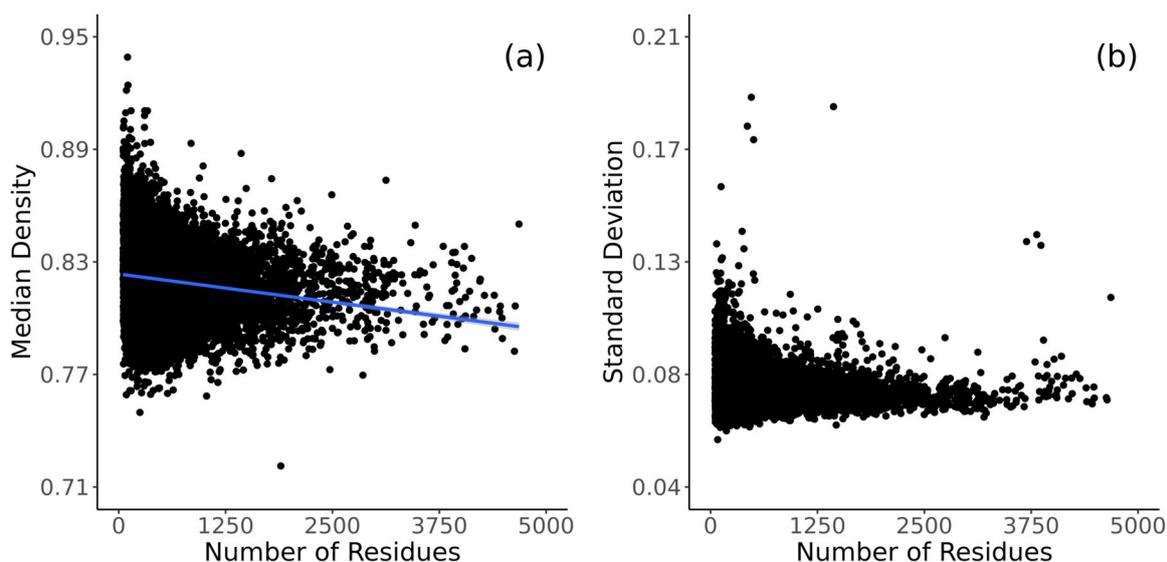

**Figure 10:** Protein size *versus* atomic density. (a) Scatter plot with regression line for number of amino acid residues *vs.* median density. The Spearman correlation is -0.18. (b) Scatter plot of the corresponding standard deviations. Notice the change of scale in the two graphs.

**Table 4 :** Median and standard deviation values in groups of structures with different size

| Size group | Median | Standard Deviation |
|---|---|---|
| 51 < aminoacids < 182 | 0.822 | 0.020 |
| 182 < aminoacids < 281 | 0.818 | 0.018 |
| 281 < aminoacids < 406 | 0.817 | 0.017 |
| 406 < aminoacids < 664 | 0.814 | 0.017 |
| 664 < aminoacids < 4682 | 0.812 | 0.016 |

Furthermore, and in terms of molecular evolution, it is expected that archaic proteins would be small size and evolutionarily actively selected to be resistant to thermal denaturation and unfolding, properties that imply the presence of increased packing density and stability. This increased stability is also known to be associated with an increased resistance to proteolysis [53,54]. With the emergence of larger proteins within the progressively more complex and compartmentalized cellular environment, the evolutionary pressure to maintain dense and stable packing was reduced, and flexible and less well-packed proteins emerged [55].

Among the rest of the factors characterizing crystallographically determined protein structures, the abundance of ordered water molecules, and the mean value of the atomic temperature factors ("B-factors") were found to be strongly correlated with atomic density. The B-factor is a measure of an atom's displacement about their mean position, and it provides a measure of the flexibility and stability of the structure [56,57]. Ordered water molecules, on the other hand, become immobilized on the surface of stable structures. In these stable structures, the reduced mobility of side chains promotes favorable interactions and bond formation with water molecules. Thus, crystallographic water abundance and B-factors can be used as indicators of stability, to examine the relationship between stability and packing. The relationship between packing density and these two factors is shown in Figures 11(c) and 11(d) respectively. Water abundance shows a strong positive correlation with median density values (Pearson +0.76, Spearman +0.78), indicating that the more densely packed a structure is, the more crystallographic waters are likely to be resolved. Conversely, the median of B-factors shows a pronounced anti-correlation with the median density (Pearson -0.56, Spearman -0.57), which indicates that the flexibility and displacements of atoms in densely packed structures are reduced, compared to loosely packed structures. This shows that packing density significantly affects structures, by providing an increased stabilization.

We have also examined the data for the presence of putative correlations between secondary structure composition and the atomic density profiles. The program STRIDE [58] was used to assign secondary structure state to each amino acid of the protein clusters identified in section §3.3 and shown in Table 3. The percentage of all secondary structure assignments was calculated across all structures of each cluster. Significant correlations were detected for β-strands and α-helices, but not for other elements, such as turns and loops. The distribution of the elements' percentages is shown in Figures 11(a) and 11(b). Examination of these figures indicates that as we move from loose (clusters 4 & 1) to dense (clusters 5 & 6) clusters, the percentage of β-strands is decreasing, while the percentage of α-helices is increasing. This is an indication that certain types of structures such as, for example, coiled-coils may appear only in the dense clusters. To investigate this, we collected the structure-related keywords from the corresponding HEADER records of the PDB files. The keyword "COILED" appeared in 6.5% of the proteins of the dense cluster, but not at all in the loosely packed cluster. We have also visually inspected the coiled-coil structures that matched the keyword with molecular graphics to validate the existence of the motif. Coiled-coils are thus found to be structures with increased packing density. The coiled-coil, a slightly twisted arrangement of two or more α-helices frequently found in fibrous proteins, was proposed by Crick in 1953 [59]. These structures have a 'knobs-into-holes' type of packing in which a hydrophobic core residue from one helix is packed in a "hole" formed by four residues of the other helix, resulting in a tight side-by-side arrangement of the hydrophobic core residues [60]. In addition, the hydrophobic core present in this motif offers stabilization in these structures [61]. Since some coiled-coils appear in fibrous proteins (e.g. keratin, myosin, kinesin) with structural and motor roles, mechanical stability is needed, in order to function properly. This stability, especially when accompanied by large internal cavities (to store substances) in polymeric coiled-coil domains, makes these motifs ideal for therapeutic applications in the form of efficient drug delivery systems. Lastly, the simplicity and structural robustness of this motif makes it ideal for many other key biological processes such as transcription and communication [62].

We extended this keyword-based procedure beyond coiled-coils, and we calculated percentages of all possible PDB-derived keywords between the five clusters shown in Table 3. We then isolated those with a differential abundance in the various clusters. The one keyword that stands-out in this procedure is the keyword "CYTOCHROME" which appeared in 5.25% of the proteins of the dense cluster, but not at all at the less densely packed clusters. Cytochromes form a diverse group of proteins with only a few features in common. They all contain protoheme IX or one of its derivatives and function in electron transport [63]. Packing density is a crucial parameter for effective electron transport [64,65]. It has been shown that even slight reductions in the distances of through-space jumps in electron transport pathways, or enhancements in atomic packing density, can significantly accelerate the rate of transfer [66]. Our data are in good agreement with those observations as indicated by the over-representation of cytochromes in the dense clusters.

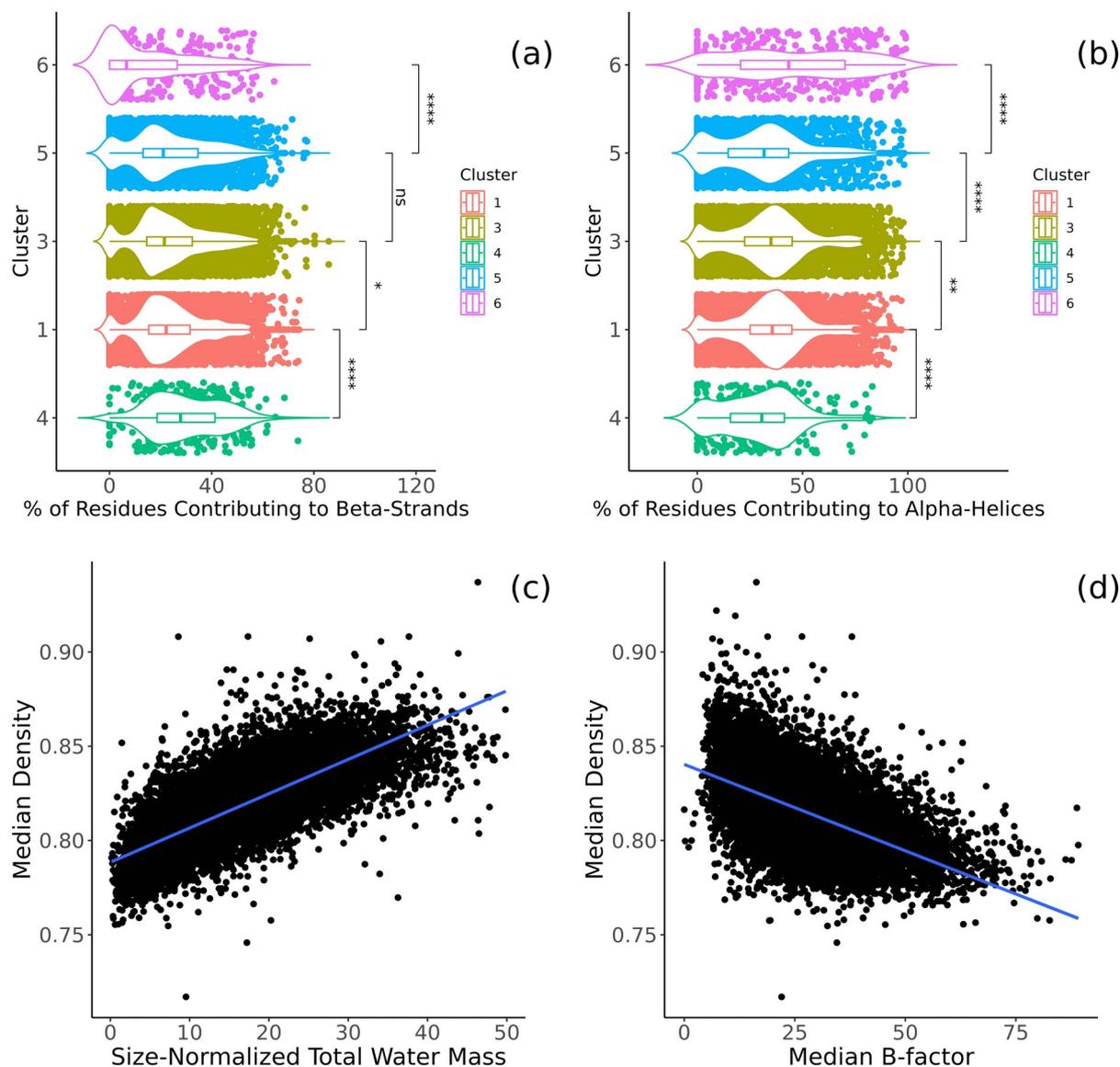

**Figure 11:** Relationship between packing density and secondary structure elements, water molecules abundance and atomic temperature factors. (a) Violin plot of the percentage of residues contributing to β-strands across clusters. The order of clusters is sorted on decreasing density (dense clusters on the top, less dense towards the bottom), (b) Violin plot of the percentage of residues contributing to α-helices across clusters. (c) Correlation between median density and abundance of water molecules. (d) Correlation between median density and the atomic temperature factors (B-factors). See text for details.

## 3.5 Functional Implications of the atomic density distributions

In this last section of our analysis we attempt to identify and characterize possible relationships between the atomic density distributions and the functional properties of the respective proteins. We aimed to explore whether certain families of proteins show a detectable preference for a loose or dense packing in order to function properly. For that purpose, we collected the classification terms from the HEADER area of each PDB file and calculated the abundances for each of the clusters we described in section §3.3. Results are shown in Figure 12.

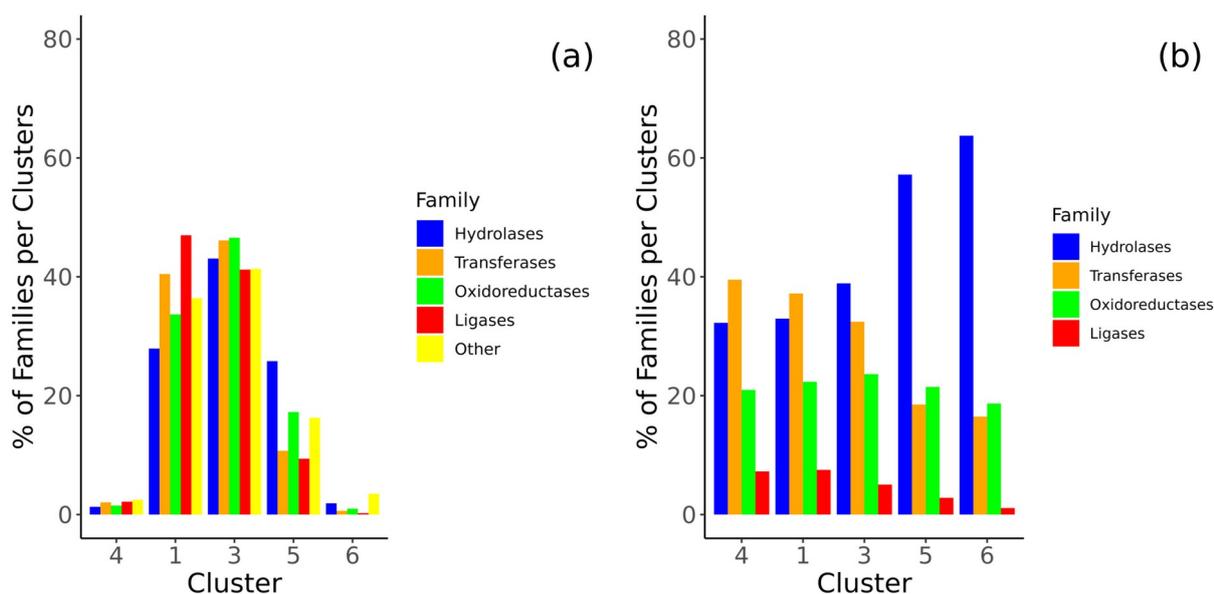

**Figure 12:** Distribution of protein families across the various clusters. In both of these diagrams, low density clusters are to the left, high density clusters to the right. Panel (a) shows how the members of a given protein family are distributed across all five clusters. The scaling is such that summing all frequencies for any given family adds-up to 100%. This diagram has not been corrected for cluster size, which explains the marked differences observed. Panel (b) shows the relative distribution of protein families on a per-cluster basis (and not across all clusters as in panel (a)).

It can be seen that the distribution of protein families across clusters is not uniform. Instead, certain groups appear in higher percentages in specific density clusters. Since cluster sizes and total family abundances vary, two different normalization approaches were applied to get comparable percentages. For Figure 12(a), the abundance of each family in a cluster was divided by the total abundance of the family across the sample of the 5 clusters (so percentages are comparable only within each cluster). We can see that Hydrolases have a higher percentage in cluster 5 (dense cluster) compared to Transferases, whereas in cluster 1 (loose cluster) the percentage of Transferases is higher than the one of Hydrolases. For Figure 12(b), the abundance of each family in a cluster was divided by the total abundance of the 4 families within this cluster and thus, family percentages are directly comparable across clusters. We can see that ligases and transferases prefer loose clusters, while hydrolases the denser ones. For oxidoreductases there is a slight preference for the average cluster (cluster 3). Furthermore, the electron transport family which was seen increased in the dense clusters does not appear in the plot. This family includes cytochromes, that were analyzed in a previous section. Other known protein families with no significant changes across clusters do not appear in the plot.

The observed differences across protein families indicate that through protein evolution, a variation in packing density occurred, and that this variation is linked with the functional properties of the respective protein families. Multiple studies have noted that the emergence of new enzymatic specificities is often linked to a decrease in the protein's thermodynamic stability, indicating the presence of a trade-off between gaining new enzymatic functions and maintaining stability [67–70]. A fine balance between stability and activity is essential for enzymes to function optimally. However, the extent of this trade-off across different protein regions and its dependence on environmental conditions remains unclear [71]. It is also important to mention that enzymes catalyzing reactions with relatively simple mechanisms (for example, hydrolysis) were likely to be one of the earliest to evolve. This, alongside the fact that earlier proteins were probably more tightly packed (see previous section), is connected with the question of whether hydrolases were among the first protein classes to be established in terms of molecular evolution. In addition, transferases, which transfer non-water functional groups, are able to prevent the mechanistically similar process of hydrolysis in the cellular environment, where water is more abundant than any other substrate [72,73]. Some transferases, which are seen to prefer a looser packing, could share a common ancestor with hydrolases especially when considering the function/stability trade-off and their similar reaction mechanism. This discussion, however, remains largely speculative and should be viewed as an open-ended question, requiring further studies on protein evolution to gain a better understanding of the underlying molecular evolution mechanisms.

## 4. Discussion

In this work, we explored the role of packing in proteins, by analyzing a large sample of crystallographically resolved, *in silico* hydrogenated and hydrated protein structures. Our results agree with and further validate previous studies on the relationships between packing and stability on one hand, and packing and protein size on the other. Small proteins are seen to have a wider variability of density values, but in general are more compact than larger ones. The reduced density of larger proteins, may result from biological constraints, as discussed in §3.4. The correlations between median density and the abundance of crystallographic water molecules and B-factors, shown in Fig. 11, supports existing knowledge concerning the effect of packing on stability. It also indicated that our algorithm for calculating atomic density distributions provides quality metrics that can be used to classify the packing level of a protein structure.

Furthermore, search for structural patterns across clusters of interest, revealed special folding patterns such as coiled-coils and cytochromes, with a high percentage in the most dense cluster. Both cases are characteristic examples of how a protein's atomic density may be connected with the functional requirements. Mechanical stability (coiled-coils) and effective electron transport (cytochromes) actively promote a denser packing which is probably needed for them to function properly.

Regarding the functional implications of atomic density, we observe a preference for some protein families on specific density clusters. This further supports the packing-function relationship we discussed for coiled-coils and cytochromes. Except for the stability aspect, which plays a crucial role in determining a protein's function, the close proximity of atoms within a structure may also be an important factor for some enzymatic reactions to happen. This is because density of packing may affect the volume and flexibility of active sites and determine which substrates have the appropriate size to be accommodated and stabilized inside them. Following this point of view may provide further insight into how transferases minimize their hydrolytic activity given that water is the most abundant substrate in the cell enrivounment : Transferases with their preference being for less well-packed structures, may be unable to interact with and stabilize water molecules. Conversely, hydrolases could immobilize water molecules more easily, as they are more stable themselves and the distances between side chains of their structures tend to be smaller.

To conclude, we have exhaustively analyzed and compared density profiles for an extended set of more than 21,000 proteins. Our analysis and subsequent all-to-all comparison of those distributions allowed us to quantitatively characterize structural and functional patterns present in these distributions, and to validate and further elaborate previous results in the field. Based on our analysis, we can corroborate the general view on the subject of protein density distributions : although systematic patterns of differences in atomic density are indeed present, these are *not* on a fixed one-to-one correspondence with structural and functional characteristics of the respective proteins. In a sense, the variability of atomic density that we observe in present-day proteins may be viewed as a remnant of the molecular evolution processes that led to those proteins, and not the direct result of a presently active selection process.

## Program and data availability

All data and programs used for the analyses reported in this communication are available *via* https://github.com/SotirisTouliopoulos/atomic_density .

# References

(1) Richards, F. M. The Interpretation of Protein Structures: Total Volume, Group Volume Distributions and Packing Density. Journal of Molecular Biology, 1974, 82, 1–14.

(2) Gregoret, L. M.; Cohen, F. E. Novel Method for the Rapid Evaluation of Packing in Protein Structures. Journal of Molecular Biology, 1990, 211, 959–974.

(3) Random Packings and the Structure of Simple Liquids. I. The Geometry of Random Close Packing. Proceedings of the Royal Society A: Mathematical, Physical and Engineering Sciences, 1970, 319, 479–493.

(4) Pontius, J.; Richelle, J.; Wodak, S. J. Deviations from Standard Atomic Volumes as a Quality Measure for Protein Crystal Structures. Journal of Molecular Biology, 1996, 264, 121–136.

(5) Gerstein, M.; Sonnhammer, E. L. L.; Chothia, C. Volume Changes in Protein Evolution. Journal of Molecular Biology, 1994, 236, 1067–1078.

(6) Ptitsyn, O. B.; Volkenstein, M. V. Protein Structures and Neutral Theory of Evolution. Journal of Biomolecular Structure and Dynamics, 1986, 4, 137–156.

(7) Lesk, A. M.; Chothia, C. How Different Amino Acid Sequences Determine Similar Protein Structures: The Structure and Evolutionary Dynamics of the Globins. Journal of Molecular Biology, 1980, 136, 225–270.

(8) Pattabiraman, N.; Ward, K. B.; Fleming, P. J. Occluded Molecular Surface: Analysis of Protein Packing. Journal of Molecular Recognition, 1995, 8, 334–344.

(9) Tanaka, S.; Scheraga, H. A. Medium- and Long-Range Interaction Parameters between Amino Acids for Predicting Three-Dimensional Structures of Proteins. Macromolecules, 1976, 9, 945–950.

(10) Warme, P. K.; Morgan, R. S. A Survey of Atomic Interactions in 21 Proteins. Journal of Molecular Biology, 1978, 118, 273–287.

(11) NARAYANA, S. V. L.; ARGOS, P. Residue Contacts in Protein Structures and Implications for Protein Folding. International Journal of Peptide and Protein Research, 1984, 24, 25–39.

(12) Miyazawa, S.; Jernigan, R. L. Residue – Residue Potentials with a Favorable Contact Pair Term and an Unfavorable High Packing Density Term, for Simulation and Threading. Journal of Molecular Biology, 1996, 256, 623–644.

(13) Abagyan, R. A.; Totrov, M. M. Contact Area Difference (CAD): A Robust Measure to Evaluate Accuracy of Protein Models. Journal of Molecular Biology, 1997, 268, 678–685.

(14) Word, J. M.; Lovell, S. C.; LaBean, T. H.; Taylor, H. C.; Zalis, M. E.; Presley, B. K.; Richardson, J. S.; Richardson, D. C. Visualizing and Quantifying Molecular Goodness-of-Fit: Small-Probe Contact Dots with Explicit Hydrogen Atoms. Journal of Molecular Biology, 1999, 285, 1711–1733.

(15) Erokhin, V.; Facci, P.; Kononenko, A.; Radicchi, G.; Nicolini, C. On the Role of Molecular Close Packing on the Protein Thermal Stability. Thin Solid Films, 1996, 284–285, 805–808.

(16) Kellis, J. T., Jr; Nyberg, K.; Sˇail, D.; Fersht, A. R. Contribution of Hydrophobic Interactions to Protein Stability. Nature, 1988, 333, 784–786.

(17) Mravic, M.; Thomaston, J. L.; Tucker, M.; Solomon, P. E.; Liu, L.; DeGrado, W. F. Packing of Apolar Side Chains Enables Accurate Design of Highly Stable Membrane Proteins. Science, 2019, 363, 1418–1423.

(18) Chen, J.; Stites, W. E. Packing Is a Key Selection Factor in the Evolution of Protein Hydrophobic Cores. Biochemistry, 2001, 40, 15280–15289.

(19) Vlassi, M.; Cesareni, G.; Kokkinidis, M. A Correlation between the Loss of Hydrophobic Core Packing Interactions and Protein Stability. Journal of Molecular Biology, 1999, 285, 817–827.

(20) Vogt, G.; Argos, P. Protein Thermal Stability: Hydrogen Bonds or Internal Packing? Folding and Design, 1997, 2, S40–S46.

(21) Sandberg, W. S.; Terwilliger, T. C. Influence of Interior Packing and Hydrophobicity on the Stability of a Protein. Science, 1989, 245, 54–57.

(22) Lim, W. A.; Sauer, R. T. The Role of Internal Packing Interactions in Determining the Structure and Stability of a Protein. Journal of Molecular Biology, 1991, 219, 359–376.

(23) Makarov, D. E.; Keller, C. A.; Plaxco, K. W.; Metiu, H. How the Folding Rate Constant of Simple, Single-Domain Proteins Depends on the Number of Native Contacts. Proceedings of the National Academy of Sciences, 2002, 99, 3535–3539.

(24) Oda, K.; Kinoshita, M. Physicochemical Origin of High Correlation between Thermal Stability of a Protein and Its Packing Efficiency: A Theoretical Study for Staphylococcal Nuclease Mutants. Biophysics and Physicobiology, 2015, 12, 1–12.

(25) Pandurangan, A. P.; Ochoa-Montaño, B.; Ascher, D. B.; Blundell, T. L. SDM: A Server for Predicting Effects of Mutations on Protein Stability. Nucleic Acids Research, 2017, 45, W229–W235.

(26) Schell, D.; Tsai, J.; Scholtz, J. M.; Pace, C. N. Hydrogen Bonding Increases Packing Density in the Protein Interior. Proteins: Structure, Function, and Bioinformatics, 2006, 63, 278–282.

(27) Sugita, Y.; Kitao, A.; Go, N. Computational Analysis of Thermal Stability: Effect of Ile→Val Mutations in Human Lysozyme. Folding and Design, 1998, 3, 173–181.

(28) Prevost, M.; Wodak, S. J.; Tidor, B.; Karplus, M. Contribution of the Hydrophobic Effect to Protein Stability: Analysis Based on Simulations of the Ile-96----Ala Mutation in Barnase. Proceedings of the National Academy of Sciences, 1991, 88, 10880–10884.

(29) Kono, H.; Saito, M.; Sarai, A. Stability Analysis for the Cavity-Filling Mutations of the Myb DNA-Binding Domain Utilizing Free-Energy Calculations. Proteins: Structure, Function, and Bioinformatics, 2000, 38, 197–209.

(30) Tsai, J.; Taylor, R.; Chothia, C.; Gerstein, M. The Packing Density in Proteins: Standard Radii and Volumes. Journal of Molecular Biology, 1999, 290, 253–266.

(31) Sonavane, S.; Chakrabarti, P. Cavities and Atomic Packing in Protein Structures and Interfaces. PLoS Computational Biology, 2008, 4, e1000188.

(32) Halle, B. Flexibility and Packing in Proteins. Proceedings of the National Academy of Sciences, 2002, 99, 1274–1279.

(33) Gerstein, M.; Tsai, J.; Levitt, M. The Volume of Atoms on the Protein Surface: Calculated from Simulation, Using Voronoi Polyhedra. Journal of Molecular Biology, 1995, 249, 955–966.

(34) Gerstein, M.; Chothia, C. Packing at the Protein-Water Interface. Proceedings of the National Academy of Sciences, 1996, 93, 10167–10172.

(35) Lim, W. A.; Sauer, R. T. Alternative Packing Arrangements in the Hydrophobic Core of λrepresser. Nature, 1989, 339, 31–36.

(36) Huang, P.-S.; Boyken, S. E.; Baker, D. The Coming of Age of de Novo Protein Design. Nature, 2016, 537, 320–327.

(37) Gekko, K. Volume and Compressibility of Proteins. Subcellular Biochemistry, 2015, 72, 75–108.

(38) Chalikian, T. V.; Gindikin, V. S.; Breslauer, K. J. Volumetric Characterizations of the Native, Molten Globule and Unfolded States of Cytochromecat Acidic pH. Journal of Molecular Biology, 1995, 250, 291–306.

(39) Perutz, M. F.; Kendrew, J. C.; Watson, H. C. Structure and Function of Haemoglobin: II. Some Relations between Polypeptide Chain Configuration and Amino Acid Sequence. Journal of Molecular Biology, 1965, 13, 669–678.

(40) Kimura, M.; Ohta, T. On Some Principles Governing Molecular Evolution. Proceedings of the National Academy of Sciences, 1974, 71, 2848–2852.

(41) Echave, J.; Spielman, S. J.; Wilke, C. O. Causes of Evolutionary Rate Variation among Protein Sites. Nature Reviews Genetics, 2016, 17, 109–121.

(42) Shih, C.; Chang, C.; Lin, Y.; Lo, W.; Hwang, J. Evolutionary Information Hidden in a Single Protein Structure. Proteins: Structure, Function, and Bioinformatics, 2012, 80, 1647–1657.

(43) Lin, C.; Huang, S.; Lai, Y.; Yen, S.; Shih, C.; Lu, C.; Huang, C.; Hwang, J. Deriving Protein Dynamical Properties from Weighted Protein Contact Number. Proteins: Structure, Function, and Bioinformatics, 2008, 72, 929–935.

(44) Lobanov, M. Yu.; Bogatyreva, N. S.; Galzitskaya, O. V. Radius of Gyration as an Indicator of Protein Structure Compactness. Molecular Biology, 2008, 42, 623–628.

(45) Fischer, H.; Polikarpov, I.; Craievich, A. F. Average Protein Density Is a Molecular-weight-dependent Function. Protein Science, 2004, 13, 2825–2828.

(46) Fleming, P. J.; Richards, F. M. Protein Packing: Dependence on Protein Size, Secondary Structure and Amino Acid Composition. Journal of Molecular Biology, 2000, 299, 487–498.

(47) Wang, G.; Dunbrack, R. L., Jr. PISCES: A Protein Sequence Culling Server. Bioinformatics, 2003, 19, 1589–1591.

(48) O'Boyle, N. M.; Banck, M.; James, C. A.; Morley, C.; Vandermeersch, T.; Hutchison, G. R. Open Babel: An Open Chemical Toolbox. Journal of Cheminformatics, 2011, 3, 33.

(49) Rodrigues, J. P. G. L. M.; Teixeira, J. M. C.; Trellet, M.; Bonvin, A. M. J. J. Pdb-Tools: A Swiss Army Knife for Molecular Structures. F1000Research, 2018, 7, 1961.

(50) Liang, J.; Dill, K. A. Are Proteins Well-Packed? Biophysical Journal, 2001, 81, 751–766.

(51) Zhang, J.; Chen, R.; Tang, C.; Liang, J. Origin of Scaling Behavior of Protein Packing Density: A Sequential Monte Carlo Study of Compact Long Chain Polymers. The Journal of Chemical Physics, 2003, 118, 6102–6109.

(52) Rawat, N.; Biswas, P. Shape, Flexibility and Packing of Proteins and Nucleic Acids in Complexes. Physical Chemistry Chemical Physics, 2011, 13, 9632-9643.

(53) Ahmad, S.; Kumar, V.; Ramanand, K. B.; Rao, N. M. Probing Protein Stability and Proteolytic Resistance by Loop Scanning: A Comprehensive Mutational Analysis. Protein Science, 2012, 21, 433–446.

(54) Daniel, R. M.; Cowan, D. A.; Morgan, H. W.; Curran, M. P. A Correlation between Protein Thermostability and Resistance to Proteolysis. Biochemical Journal, 1982, 207, 641–644.

(55) Tang, Q.-Y.; Ren, W.; Wang, J.; Kaneko, K. The Statistical Trends of Protein Evolution: A Lesson from AlphaFold Database. Molecular Biology and Evolution, 2022, 39.

(56) Frauenfelder, H.; Sligar, S. G.; Wolynes, P. G. The Energy Landscapes and Motions of Proteins. Science, 1991, 254, 1598–1603.

(57) Sun, Z.; Liu, Q.; Qu, G.; Feng, Y.; Reetz, M. T. Utility of B-Factors in Protein Science: Interpreting Rigidity, Flexibility, and Internal Motion and Engineering Thermostability. Chemical Reviews, 2019, 119, 1626–1665.

(58) Heinig, M.; Frishman, D. STRIDE: A Web Server for Secondary Structure Assignment from Known Atomic Coordinates of Proteins. Nucleic Acids Research, 2004, 32, W500–W502.

(59) Crick, F. H. C. The Packing of α-Helices: Simple Coiled-Coils. Acta Crystallographica, 1953, 6, 689–697.

(60) Lupas, A. N.; Gruber, M. The Structure of α-Helical Coiled Coils. Advances in Protein Chemistry, 2005, 70, 37–38.

(61) Yu, Y. B. Coiled-Coils: Stability, Specificity, and Drug Delivery Potential. Advanced Drug Delivery Reviews, 2002, 54, 1113–1129.

(62) McFarlane, A. A.; Orriss, G. L.; Stetefeld, J. The Use of Coiled-Coil Proteins in Drug Delivery Systems. European Journal of Pharmacology, 2009, 625, 101–107.

(63) Scott Mathews, F. The Structure, Function and Evolution of Cytochromes. Progress in Biophysics and Molecular Biology, 1985, 45, 1–56.

(64) Nowak, C.; Schach, D.; Gebert, J.; Grosserueschkamp, M.; Gennis, R. B.; Ferguson-Miller, S.; Knoll, W.; Walz, D.; Naumann, R. L. C. Oriented Immobilization and Electron Transfer to the Cytochrome c Oxidase. Journal of Solid State Electrochemistry, 2011, 15, 105–114.

(65) Shao, H.-B.; Yu, H.-Z.; Zhao, J.-W.; Zhang, H.-L.; Liu, Z.-F. The Effect of Packing Density on the Electron-Transfer Kinetics of an Azobenzenethiol Monolayer on Gold. Chemistry Letters, 1997, 26, 749–750.

(66) Williamson, H. R.; Dow, B. A.; Davidson, V. L. Mechanisms for Control of Biological Electron Transfer Reactions. Bioorganic Chemistry, 2014, 57, 213–221.

(67) Tokuriki, N.; Stricher, F.; Serrano, L.; Tawfik, D. S. How Protein Stability and New Functions Trade Off. PLoS Computational Biology, 2008, 4, e1000002.

(68) Studer, R. A.; Christin, P.-A.; Williams, M. A.; Orengo, C. A. Stability-Activity Tradeoffs Constrain the Adaptive Evolution of RubisCO. Proceedings of the National Academy of Sciences, 2014, 111, 2223–2228.

(69) DePristo, M. A.; Weinreich, D. M.; Hartl, D. L. Missense Meanderings in Sequence Space: A Biophysical View of Protein Evolution. Nature Reviews Genetics, 2005, 6, 678–687.

(70) Tokuriki, N.; Tawfik, D. S. Stability Effects of Mutations and Protein Evolvability. Current Opinion in Structural Biology, 2009, 19, 596–604.

(71) Hou, Q.; Rooman, M.; Pucci, F. Enzyme Stability-Activity Trade-Off: New Insights from Protein Stability Weaknesses and Evolutionary Conservation. Journal of Chemical Theory and Computation, 2023, 19, 3664–3671.

(72) Koshland, D. E. Enzyme Flexibility and Enzyme Action. Journal of Cellular and Comparative Physiology, 1959, 54, 245–258.

(73) Light, S. H.; Cahoon, L. A.; Mahasenan, K. V.; Lee, M.; Boggess, B.; Halavaty, A. S.; Mobashery, S.; Freitag, N. E.; Anderson, W. F. Transferase Versus Hydrolase: The Role of Conformational Flexibility in Reaction Specificity. Structure, 2017, 25, 295–304.